\documentclass{iau} 
\usepackage{graphicx}
\usepackage{wrapfig}

\newcommand{\sfe}{$\mathrm{SFE_{gl}}$}

\title[Star Clusters in the Galactic tidal field] 
{Star Clusters in the Galactic tidal field, \\from birth to dissolution}

\author[B. Shukirgaliyev, G. Parmentier, P. Berczik and A. Just]   
{{Bekdaulet Shukirgaliyev}$^{1,2}$, Genevieve Parmentier$^{3}$, Peter Berczik$^{4,3,5}$ \and Andreas Just$^{3}$}

\affiliation{
$^{1}$Energetic Cosmos Laboratory, Nazarbayev University, \\53 Kabanbay batyr ave., 010000 Nur-Sultan, Kazakhstan\\ email: {\tt bekdaulet.shukirgaliyev@nu.edu.kz} \\[\affilskip]
$^{2}$Fesenkov Astrophysical Institute, \\ 23 Observatory str., 050020 Almaty, Kazakhstan\\[\affilskip] 
$^{3}$Astronomisches Rechen-Institut, Zentrum f\"ur Astronomie der Universit\"at Heidelberg, \\M\"onchhofstr 12-14, D-69120 Heidelberg, Germany\\[\affilskip] 
$^{4}$National Observatories of China, Chinese Academy of Science, \\20A Datun Rd, Chaoyang District, 100012 Beijing, PR China\\[\affilskip]
$^{5}$Main Astronomical Observatory, Academy of Science of Ukraine, \\27 Akademika Zabolotnoho St, 03680 Kyiv, Ukraine 
} 

\pubyear{2019}
\volume{351}  
\jname{Star Clusters: From the Milky Way to the Early Universe}
\editors{A. Bragaglia, M.B. Davies, A. Sills \& E. Vesperini, eds.}
\begin{document}

\maketitle

\begin{abstract}
We study the evolution of star clusters in the Galactic tidal field starting from their birth in molecular clumps. Our model clusters form according to the local-density-driven cluster formation model in which the stellar density profile is steeper than that of gas. As a result, clusters resist the gas expulsion better than predicted by earlier models.

We vary the impact of the Galactic tidal field $\lambda$, considering different Galactocentric distances (3-18 kpc), as well as different cluster sizes. Our model clusters survive the gas expulsion independent of $\lambda$.

We investigated the relation between the cluster mass at the onset of secular evolution and their dissolution time. The model clusters formed with a high star-formation efficiency (SFE) follow a tight mass-dependent dissolution relation, in agreement with previous theoretical studies. However, the low-SFE models present a shallower mass-dependent relation than high-SFE clusters, and most dissolve before reaching 1 Gyr (cluster teenage mortality).


\end{abstract}

\firstsection 
\section{Introduction}

Star clusters (SC) form in collapsing cold dense clumps formed out of the turbulent gas of molecular clouds (\cite[Krumholz et al. 2019]{Krumholz+19}). 
Once thermonuclear reactions begin within massive stars ($>8M_\odot$), they start to drive effectively the natal gas out of the star-forming region. Their radiation pressure, strong winds, ionizing radiation and eventually explosion as supernovae type II are enough to clear up the cluster from all its gas within a few million years (\cite[Rahner et al. 2019]{Rahner+19}, \cite[Li et al. 2019]{2019MNRAS.487..364L}, \cite[Wall et al. 2019]{2019arXiv190101132W}). 
SCs older than 5 Myr are observed to be gas-free (\cite[Leisawitz et al. 1989]{1989ApJS...70..731L}).
The star-formation efficiency (SFE), that is the mass fraction of star-forming gas converted into stars, is usually below 30\% in star-forming regions of the solar vicinity (\cite[Lada \& Lada 2003]{LL03}, \cite[Higuchi et al. 2009]{Higuch+09}). 
The weakening of the cluster gravitational potential due to the residual gas expulsion drives the SC out of virial equilibrium, possibly causing its expansion depending on the cluster dynamical state during the gas embedded phase (see e.g. \cite[Baumgardt \& Kroupa 2007]{BK07}). During their violent relaxation, SCs can lose their stars, modify their structures and masses and even dissolve without reaching a new virial equilibrium (\cite[Shukirgaliyev et al. 2017]{Bek+17}, \cite[Shukirgaliyev et al. 2018]{Bek+18}
). 
Those SCs surviving the gas expulsion are observed as stable dynamical systems at the end of violent relaxation. However, they are going to be disrupted by several dissolution mechanisms like two-body relaxation driven evaporation, stellar evolutionary mass-loss and tidal stripping by the host galaxy (e.g \cite[Baumgardt \& Makino 2003]{BM03}, \cite[Renaud 2018]{Reanud18}).

Due to the complexity of the problem, each phase of a SC life has usually been considered separately in the framework of one particular model. These different models and approaches are not always connected with each other.
For instance the detailed hydrodynamical simulations of SC formation usually cover the early evolution only and they cannot cover the full parameter space of cluster initial conditions due to their computational cost. The detailed $N$-body simulations of the long-term evolution of SCs usually start when clusters are in virial equilibrium, neglecting the violent relaxation phase and, thus, do not bear the information about the birth conditions of SCs.
 It is worth building a self-consistent model describing the full life of SCs starting from their birth until full dissolution in the tidal field of their host galaxy. Then we will be able to see a whole picture of SC evolution and to test our models by comparing them with observations. 

There are a few projects considering the evolution of SCs from formation till dissolution in cosmological simulations (e.g. \cite[Pfeffer et al. 2018]{2018MNRAS.475.4309P}, \cite[Li et al. 2017]{2017ApJ...834...69L}). Additionally to them, in this work, we propose a new approach to build a comprehensive model of the life of SCs, including a more detailed picture of their internal dynamics. 

\section{Our model}

Instead of performing hydrodynamical simulations, we assume that our SCs form according the semi-analytical local-density-driven cluster formation model of \cite[Parmentier \& Pfalzner (2013)]{PP13}. That is, stars form in centrally concentrated, spherically symmetric clump with a constant SFE per free-fall time. Therefore, as it was observed by \cite[Gutermuth et al. (2011)]{2011ApJ...739...84G}, stars are more concentrated to the center of the cluster-forming clump than the gas (see e.g. Fig.~1 in \cite[Parmentier \& Pfalzner (2013)]{PP13} and Fig.~2 in \cite[Shukirgaliyev et al. 2017]{Bek+17}). 
We use the adapted program \textsc{mkhalo} program from \texttt{falcON} package (McMillan \& Dehnen 2007) to generate stellar systems in virial equilibrium within the residual gas potential (\cite[Shukirgaliyev et al. 2017]{Bek+17}).  
Then, assuming that gas expulsion is instantaneous we perform the direct $N$-body simulations of gas-free SCs orbiting in the Galactic disk plane on circular orbits.
The parallelized high-precision direct $N$-body code phi-GRAPE/GPU (\cite[Berczik et al., 2013]{PhiGRAPE13}) is used for our simulations.
 
We consider not only the short-term violent relaxation, but also follow the long-term evolution (aka secular evolution) of our model clusters until they dissolve in the Galactic tidal field. This way we are able to cover a large parameter space with high precision.


\section{Results}

\begin{wrapfigure}{r}{0.4\textwidth}
\begin{center}
 \includegraphics[width=0.4\textwidth]{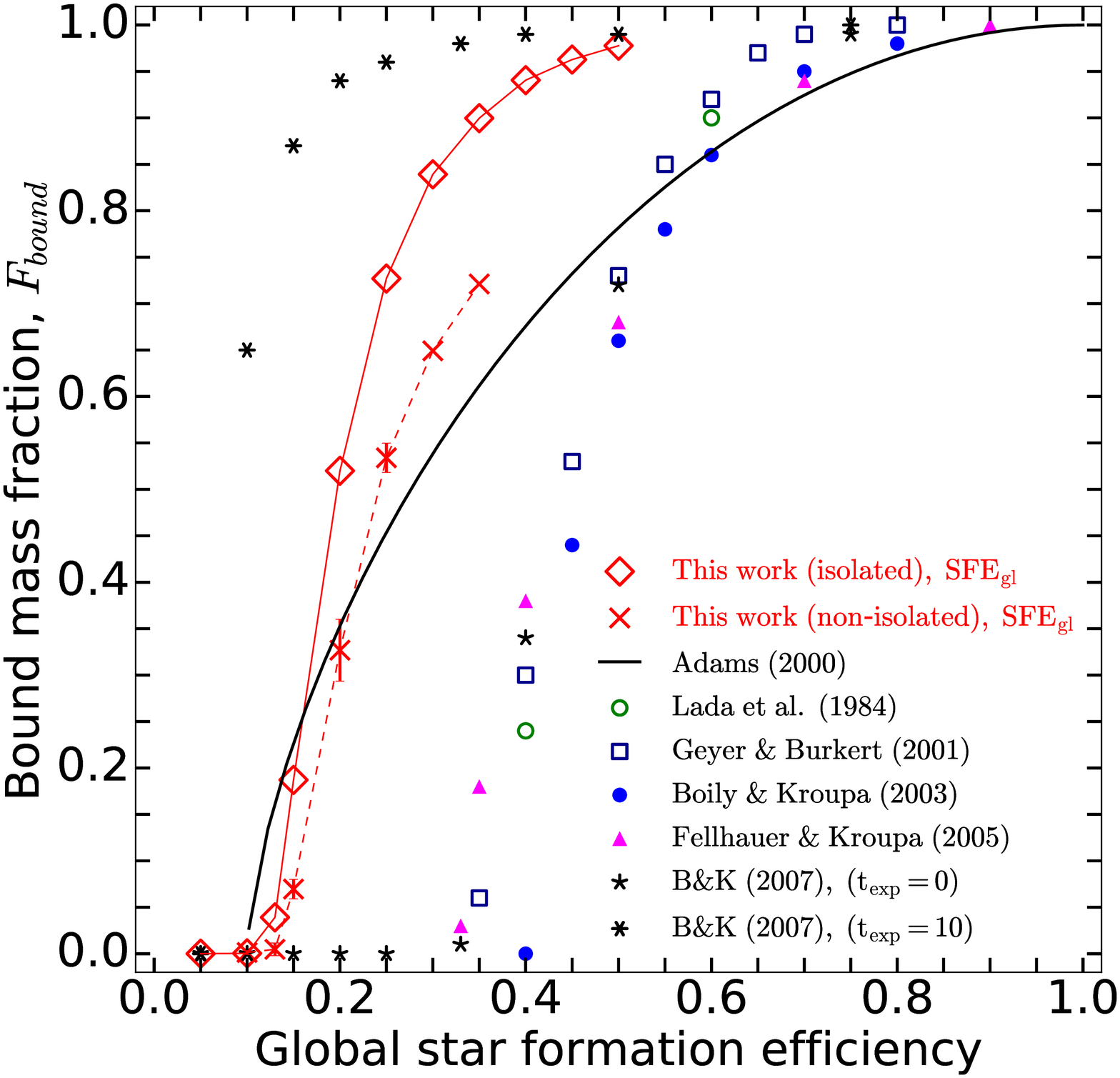} 
 \caption{Bound fraction as a function of \sfe. We compare our results (red lines) with previous works. The isolated models are depicted by red diamonds, and non-isolated models by red crosses. (Top panel of Fig.~8 in \cite[Shukirgaliyev et al. 2017]{Bek+17}).}
   \label{fig1}
\end{center}
\end{wrapfigure}

We perform $N$-body simulations of SCs evolving in the solar neighborhood (i.e. at the Galactocentric distance $R_\mathrm{orb}=8.0$~kpc) covering different global SFEs ($\mathrm{SFE_{gl}}$, the total SFE measured within dense gas clumps) from 0.1 to 0.25 (0.35) and different cluster birth masses ($M_\star$, the cluster stellar mass before gas expulsion) from 3k to 300k $\mathcal{M}_\odot$. 
We find that the violent relaxation duration (i.e. the time span over which SCs lose mass in response to gas expulsion) of our model clusters depends neither on the cluster birth mass 
nor on the \sfe.  
We find that our model clusters can survive instantaneous gas expulsion with a minimum \sfe\ of 15\% (Fig.~\ref{fig1}). 
This is a factor of two lower than previous estimates of 33\% (\cite[Shukirgaliyev et al. 2017]{Bek+17}). 
Figure~\ref{fig1} presents the bound mass fraction of model SCs at the end of violent relaxation as a function of \sfe, where we also compare our results with previous works. We have improved the survival likelihood of SCs after instantaneous gas expulsion. This is caused by the difference in density profiles between the embedded cluster and its residual gas, namely, the stellar density profile has a steeper slope than that of the residual gas (see \cite[Parmentier \& Pfalzner 2013]{PP13} for details). 
We also report that there is no dependency between $M_\star$ and the final bound fraction in our models (see Fig.~3 of \cite[Shukirgaliyev et al. 2018]{Bek+18}).

\begin{figure}[h]
\begin{center}
 \includegraphics[width=0.41\textwidth]{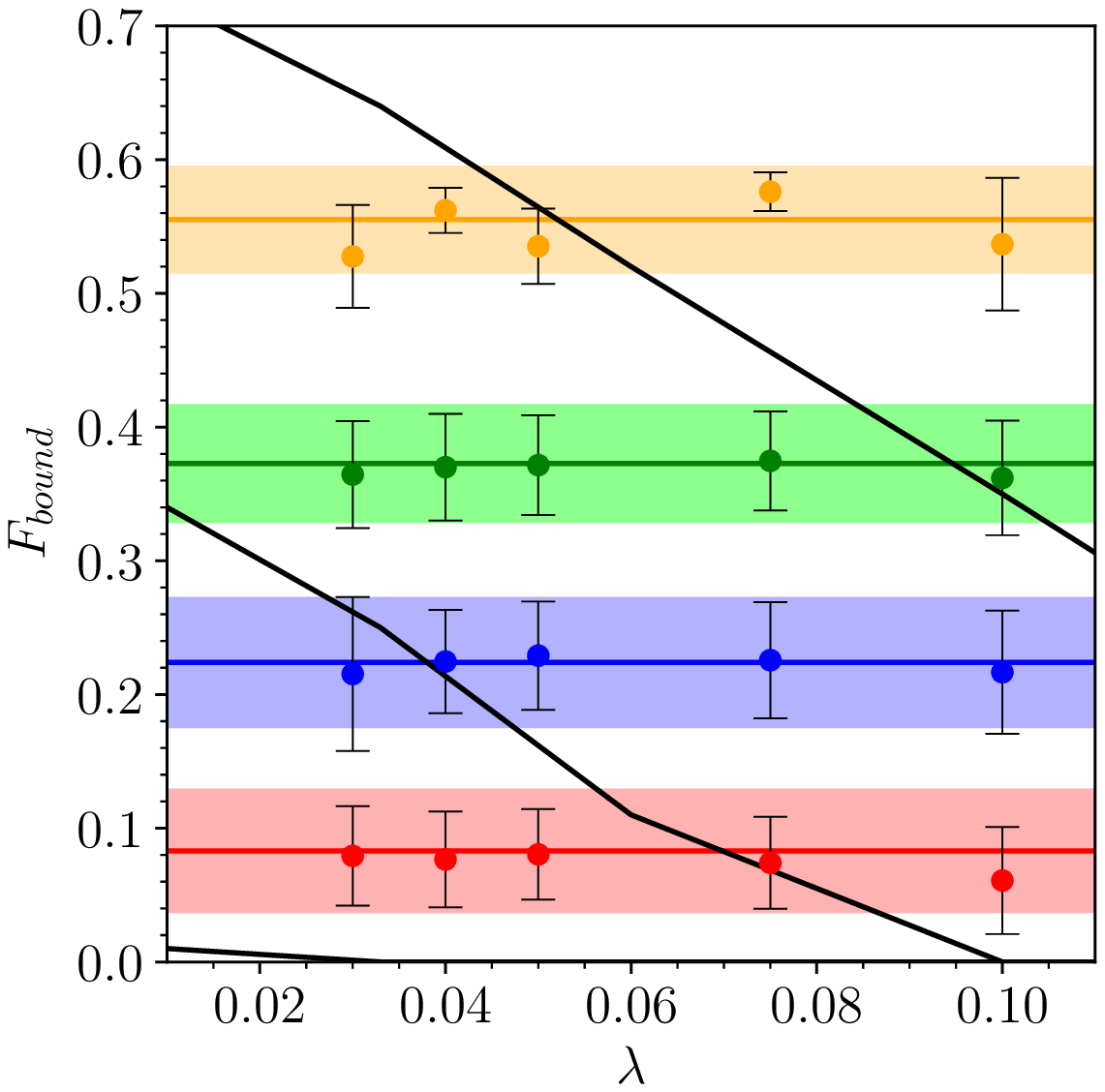}  \includegraphics[width=0.45\textwidth]{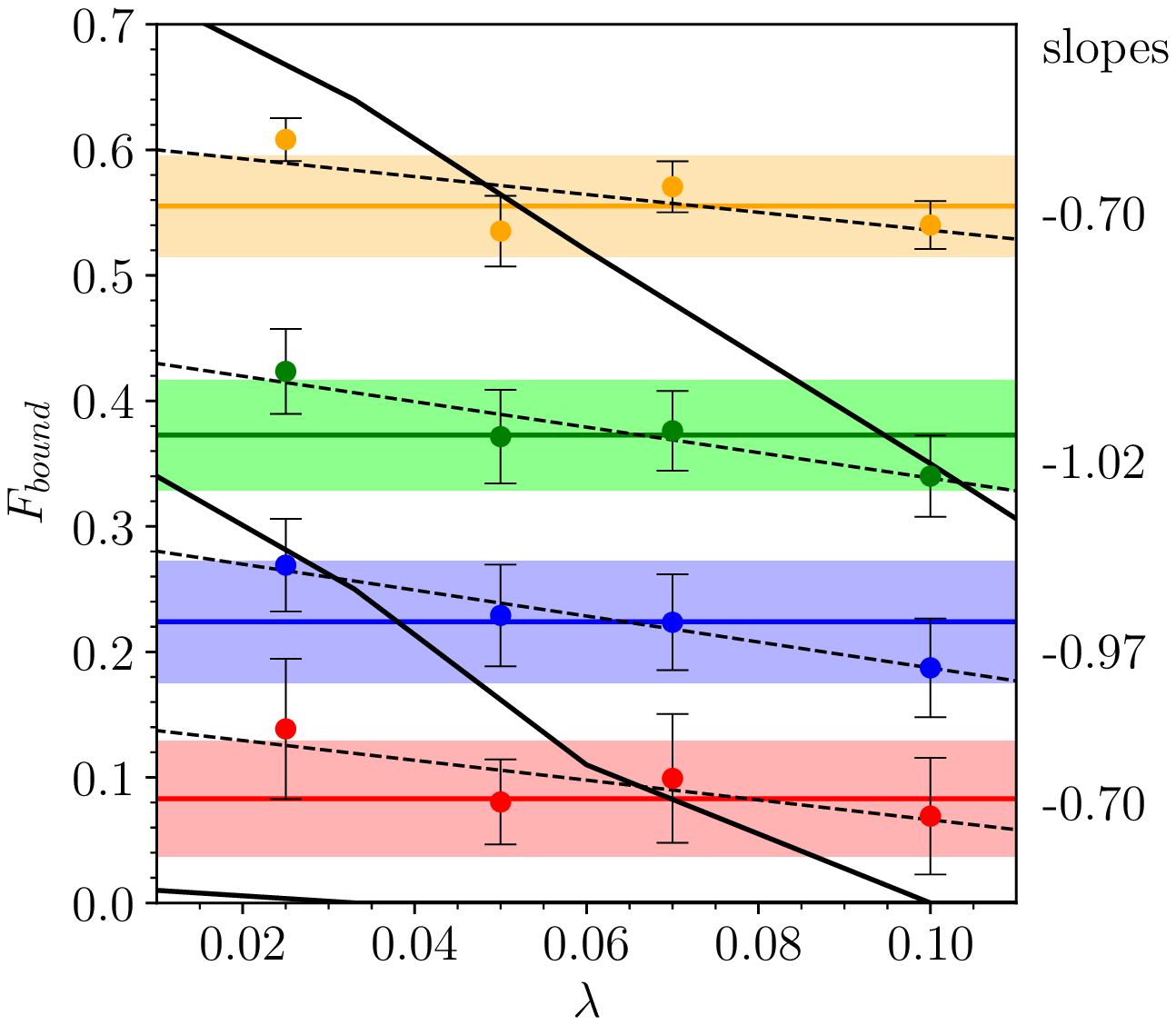} 
 \caption{The final bound mass fraction as a function of the impact of the tidal field for different Galactocentric distances (left) and different cluster central densities at $R_\mathrm{orb}~=~8.0~\mathrm[kpc]$ (right panel). The colors code $\mathrm{SFE_{gl}} = 0.15$ (red), 0.17 (blue), 0.20 (green), and 0.25 (yellow). Each point corresponds to the mean and standard deviation of model clusters with the same \sfe\ and $\lambda$. Colored solid lines and shaded areas show the mean bound fraction and standard deviation for all models with a given \sfe. 
 The black dashed lines are the linear fits to the final bound fractions of a given \sfe\ as a function of $\lambda$.
 The black solid lines correspond to the results of \cite[Baumgard \& Kroupa (2007)]{BK07} for models with an SFEs of 0.33, 0.4 and 0.5 from bottom to top, respectively. (Based on Figs.~3-5 of \cite[Shukirgaliyev et al. 2019]{Bek+19})}
   \label{fig2}
\end{center}
\end{figure}

We extend our simulations by varying the impact of the Galactic tidal field, $\lambda=r_\mathrm{h}/R_\mathrm{J}$, measured by the ratio between the half-mass radius and the Jacobi radius of clusters immediately after gas expulsion (\cite[Shukirgaliyev et al. 2019]{Bek+19}). We vary $\lambda$ in two ways: either by varying the Galactocentric distance of SCs (and therefore the Jacobi radius), by varying the radius of SCs while keeping them in the solar neighborhood ($R_\mathrm{orb}=8.0$~kpc). In the first case, we adopt $R_\mathrm{orb} = 2.9, 4.64, 8.0, 10.95, 18.0$~kpc, yielding $\lambda=[0.1,0.07,0.05,0.04,0.03]$ (left panel of Fig.~\ref{fig2}). In the second case we adopt $\lambda=[0.1,0.075,0.05,0.025]$ (right panel of Fig.~\ref{fig2}). 
Figure~\ref{fig2} shows the bound mass fraction of our model clusters at the end of violent relaxation as a function of $\lambda$. The dependency seen in the right panel of Fig.~\ref{fig2} is weak and stays within the error-bars. Therefore, we conclude that the survivability after instantaneous gas expulsion of SCs formed with a centrally peaked SFE profile is independent of the impact of the tidal field of the Galaxy (Fig.~\ref{fig2}, \cite[Shukirgaliyev et al. 2019]{Bek+19}).
 
In \cite[Shukirgaliyev et al. (2018)]{Bek+18} we consider the long-term evolution of model solar neighborhood SCs, continuing our simulations until full their dissolution in the Galactic tidal field. 
We have found that SCs formed with a high \sfe\ ($\geq0.20$) dissolve in a tight mass-dependent dissolution time (MDD) regime, in agreement with earlier works.
 In contrast, model clusters formed with low \sfe\ ($=0.15$) dissolve more quickly than high-SFE clusters, most dissolving within 1 Gyr, and present a shallower MDD relation (see Figs.~5-7 in \cite[Shukirgaliyev et al. 2018]{Bek+18}).
We find that the relation between the SC dissolution time and ``initial'' mass (i.e. cluster mass after violent relaxation, initial for long-term evolution) becomes close to that observationally found by \cite[Lamers et al. (2005)]{Lamers+05a} for the solar neighborhood if the SC population is dominated by low-SFE ones.
 Therefore, in this study, we propose to  {recover the cluster dissolution time for the solar neighborhood found by \cite[Lamers et al. (2005)]{Lamers+05a} in a way that is alternative} to that proposed by \cite[Gieles et al. (2006)]{Gieles+06}.

\vspace{1em}

 \underline{Acknowledgements.} {\footnotesize This work was supported by Sonderforschungsbereich SFB 881 ‘The Milky Way System’ (subproject B2) of the German Research Foundation (DFG). B.S. gratefully acknowledges the conference organizers for partial support in the form of IAU travel grant, the support within PCF-program BR05236322 and grant AP05135753 funded by the Ministry of 
 Education and Science of the Republic of Kazakhstan. 
}


\end{document}